\documentclass[aps,prl,showpacs,twocolumn]{revtex4}
\usepackage{epsfig,graphicx}

\input{tcilatex}

\begin{document}

\title{An Important Properties of Entanglement: Pairwise Entanglement can
Only be Transferred by Entangled Pair}
\author{Xiao-Qiang Xi$^{1,2}$, W. M. Liu$^2$}
\address{$^1$Department of Applied Mathematics and Physics,
 Xi'an Institute of Posts and Telecommunications, Xi'an 710061, China}
\address{$^2$Beijing National Laboratory for Condensed Matter Physics,
Institute of Physics, Chinese Academy of Sciences, Beijing 100080,
China}

\date{\today}

\begin{abstract}
Basing on the calculation of all the pairwise entanglement in the
$n$ ($n \leq 6$)-qubit Heisenberg XX open chain with system
impurity, we find an important result: pairwise entanglement can
only be transferred through entangled pair. The non-nearest pairwise
entanglement will has the possibility to exist as long as there
has even number qubit in their middle. This point means that we
can get longer distance entanglement in solid system.
\end{abstract}


\maketitle

It is well known that entanglement has some prominent
applications: firstly, it can be used to test some fundamental
questions of the quantum mechanics; secondly, it is the key
ingredient of the quantum information processing such as quantum
teleportation, superdense cording, quantum computation, quantum
communication, quantum computational speed-ups and quantum
cryptographic protocols; thirdly, the entangled states can be used
to the sensitivity of interferometric measurements such as quantum
lithography \cite{BotoPRL85_2733}, quantum optical gyroscope
\cite{DowlingPRA57_4736}, quantum clock synchronization and
positioning \cite{JozsaPRL85_2010} and frequency metrology
\cite{BollingerPRA54R4649}; fourthly is that entanglement play a
central role in the study of strongly correlated quantum systems
\cite{PreskillJMO47_127}, ground-state entanglement is help us to
understand the quantum phase transition
\cite{OsbornePRA66_032110}, Mott insulator-superfluid transition
and quantum magnet-paramagnet transition.

In all the implications of entanglement, the most important thing
is to find entangled pair. The photons \cite{Bouwmeester97N390},
the energy level of different trapped ions \cite{Turchette98PRL81}
(or different atoms \cite{Rauschenbeutel00S288}), qubit in crystal
lattices \cite{Yamaguchi99APA68}, qubit in Josephson junctions
\cite{Makhlin99N398} and Bose-Einstein condensates
\cite{Sorensen01N409} are often used to produce entanglement, but
their complicated equipment will prevent them from using
cosmically. While using spin chains \cite{Gershenfeld97S275} to
produce entanglement seems more convenient, because it can be
scaled easily and its equipment will be very simple, many
interesting work are developed around the entanglement in spin
chain(see Ref. \cite{Arnesen2000} and their references), while
most of them \cite{WangPRA66_034302,XiPLA300_567} are concerned
with the nearest pairwise entanglement, that is important but far
from perfect. The best entanglement should be the non-nearest
pairwise, for example, in a three qubits Heisenberg XX open chain,
the entanglement between the first and the third qubits (if exist)
is more practical than the first and the second because the former
has farther entanglement distance.

If we can find the law of the non-nearest pairwise entanglement in
spin chain, then we have the possibility to construct a longer
distance entangled pair, this is one aim of this paper. We chose
the Heisenberg XX open chain as the studying object, because it is
simple and very useful in quantum information processing, such as
it can be used for quantum computation
\cite{Lidar1999,Divincenzo2000,SantosPRA67_062306,Imamoglu1999}
and quantum communication
\cite{PRL91_207901,PRA69_034304,PRA69_052315}.

 In recently, the non-nearest pairwise entanglement in Heisenberg
 chain is also discussed
\cite{CaoPRA71_034311,XiAPS55_3026}, the non-nearest entanglement
comes from the introduce of magnetic field, which give the direct
interaction between the non-nearest spin qubits, but they have two
disadvantages, the first is its entanglement cannot be get the
maximal value 1(the numerical results
\cite{CaoPRA71_034311,XiAPS55_3026} show that the non-nearest
entanglement does not exceed 0.5), because the magnetic field only
give the interaction between the z component of the non-nearest
spin qubits; the second is that magnetic field will take
complexness to the system and its application.

So we want to seek the relation between the pairwise entanglement
and the system impurity, obviously, the system impurity is simpler
than magnetic impurity, and maybe we can find the maximal
non-nearest pairwise entanglement. This idea comes from our
previous paper \cite{XiPLA297_291}, in which we have introduced
system impurity to three qubit Heisenberg XX spin ring and find
that system impurity can control the nearest pairwise
entanglement, so we suppose that the system impurity will help us
to understand the non-nearest pairwise entanglement.

The main tool of discussing pairwise entanglement is the concept
of entanglement of formation (EoF) and concurrence
\cite{Wootters1998,Hill1997}, EoF can be used to measure the
pairwise entanglement, concurrence ${C}$ range from zero to one
and it is monotonically relate to EoF, so concurrence is often
used to measure the pairwise entanglement, we adopt this
measurement in this paper and use $C_{ij}$ to express the pairwise
entanglement between the $i$th and $j$th qubits. ${C}_{ij}=
\mbox{max} \{ {\lambda}_{1} -{\lambda}_{2} -{\lambda}_{3}
-{\lambda}_{4},0 \}$, where $\lambda_k$, $k=1,2,3,4$ are the
square roots of the eigenvalues of the operator
${\hat{\rho}}_{ij}= {\rho}_{ij}
({\sigma}_i^y{\otimes}{\sigma}_j^y) {\rho}_{ij}^*
({\sigma}_i^y{\otimes}{\sigma}_j^y)$ in descending order,
${\rho}_{ij}= Tr_{non(ij)} \rho$ is the reduced density matrix of
the system.

In order to introduce the impurity, we write the Hamiltonian of
the N-qubit Heisenberg XX open chain as following
\begin{equation}
H= \sum_{i=1}^{N-1} J_{i} J_{i+1} (\sigma_{i}^{+} \sigma_{i+1}^{-}
+\sigma_{i+1}^{+} \sigma_{i}^{-}), \label{Ham}
\end{equation}
where $\sigma^{\pm}= \frac{1}{2} (\sigma^x \pm \sigma^y)$,
$\sigma^{x}, \sigma^{y}, \sigma^{z}$ are the Pauli matrices, $J_i$
is the contribution of the $i$th lattice to the exchange hopping
and $J_{i}J_{i+1}$ is the exchange hopping between the $i$th and
the $(i+1)$th lattice. If the $i$th lattice is impurity, then we
let $J_{i}$ as the impurity parameter, $J_{k}=1$ $(k \ne i)$ are
the normal parameters.

First we begin from the simplest case of the above Hamiltonian,
i.e $N=3$, different impurity can be seen from Fig. 1. The results
of calculation show that $J_1$ or $J_3$ can make $C_{12}=1$, $J_2$
can only make $(C_{12})_{max}=0.457$, those can be understand
easily, increasing $J_1$ or decreasing $J_3$ will couple stronger
the first and the second qubits, i.e increasing their interaction
and entanglement, while $J_2$ is an overall factor, in this sense
this impurity system is equal to a normal system. Another main
task of us is to discuss the relation between $C_{13}$ and
impurity parameter $J_i$, unfortunately, $C_{13}=0$ no matter how
we change the impurity parameter, which seems no confusion because
there is no direct interaction between them.

\begin{figure}[h]
\center{\epsfxsize 3cm \epsfysize 2cm \epsffile{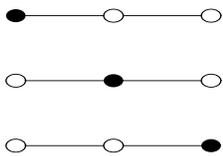}}
\caption{The diagram of the impurity position in the 3-qubit open
chain, where the solid circle is the impurity qubit, the hollow
circle are the normal qubits. The site are labelled 1,2,3 from
left to right.}
\end{figure}

Then let us see the next simplest case, i.e $N=4$. As the impurity
changing from the first to the forth site, we need to discuss the
value of $C_{12},C_{23},C_{13}$ and $C_{14}$. $C_{12}$ and
$C_{23}$ are the nearest pairwise entanglement, $C_{13}$ and
$C_{24}$ are the non-nearest pairwise entanglement. Different
impurity can be seen from Fig. 2.

\begin{figure}[h]
\center{\epsfxsize 3cm \epsfysize 2cm \epsffile{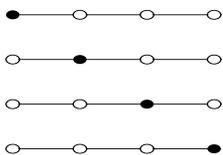}}
\caption{The diagram of the impurity position in the four-qubit
open chain, where the solid circle is the impurity qubit, the
hollow circle are the normal qubits. The site are labelled 1,2,3,4
from left to right.}
\end{figure}

Tedious calculations show the following results:

(1) Changing every impurity site, increasing $J_1$ (or $J_4$) and
decreasing $J_2$ (or $J_3$), can make $C_{12}=1$, the physical
meaning are clear: increasing $J_1$ (or $J_4$) and decreasing
$J_2$ (or $J_3$) means that we isolate the first and the second
qubits from the whole chain, their interaction is equal to the two
qubit case, of cause it can get its maximal entanglement. (2) The
maximal value of $C_{23}$ is $0.457$, the physical nature lie when
increasing $J_1$ (or $J_4$) or decreasing $J_2$ (or $J_3$) means
that we couple weaker the middle qubits, of course will decrease
their entanglement; decreasing $J_1$(or $J_4$) or increasing $J_2$
(or $J_3$) means that we will take a qubit away from the whole
chain, so the four-qubit chain become a three-qubit chain with
uniform coupling, its maximal value is just $0.457$. (3)
$C_{13}=0$ no matter how we change the impurity parameters, i.e
there is no entanglement between the next nearest qubits, we still
consider that there is no direct interaction between them. (4)
There exist entanglement between the boundary qubits, its maximal
value $(C_{14})_{max}=0.457$, $C_{14}$ increases as we decrease
$J_1$ (or $J_4$) or increase $J_2$ (or $J_3$). Obviously, there is
no direct interaction between the boundary quabits. No direct
interaction theory seems invalidation. Certainly there exist
another reason, which is just what we seek for.

No direct interaction theory can not give a harmonious explanation
between $C_{13}=0$ and $C_{14} \ne 0$, is there some different
between them? Let us see this question between three-qubit and
four-qubit cases, See Fig. 3.

\begin{figure}[h]
\center{\epsfxsize 6cm \epsfysize 4cm \epsffile{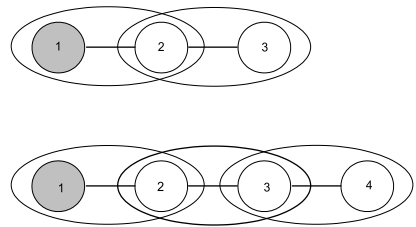}}
\caption{The diagram of the comparison three-qubit and four-qubit
open chain, where the solid circle is the impurity qubit, the
hollow circle are the normal qubits, the ellipse denotes that
there exist entanglement between the qubits in it.}
\end{figure}

The different is very clearly: there is only one qubit in the
middle of the first and the third qubit, while two qubits (an
entangled pair) in the first and the forth qubit. Obviously, the
middle qubit(s) plays a key role in this different.

We suppose such an explanation: for three-qubit case, there is
only one qubit in the middle, one qubit can not entangle with
itself, although $C_{12}\ne 0$ and $C_{23}\ne 0$, $C_{13}=0$, as
if the entanglement is broken in the middle; while for four-qubit
case, there is an entangled pair in the middle, $C_{12} \ne 0,
C_{23} \ne 0$ and $C_{34} \ne 0$, so $C_{14}\ne 0$, i.e
entanglement can be transferred through an entangled pair.

For understanding more details about $C_{14}$ we give a comparison
among $C_{12}, C_{23}, C_{34}$ and $C_{14}$ through plotting the
figures $C_{ij}$ at fixed $J_1=J$ (the first site is impurity) and
fixed temperature, see Fig. 4.

\begin{figure}[h]
{\epsfxsize 3cm \epsfysize 3cm \epsffile{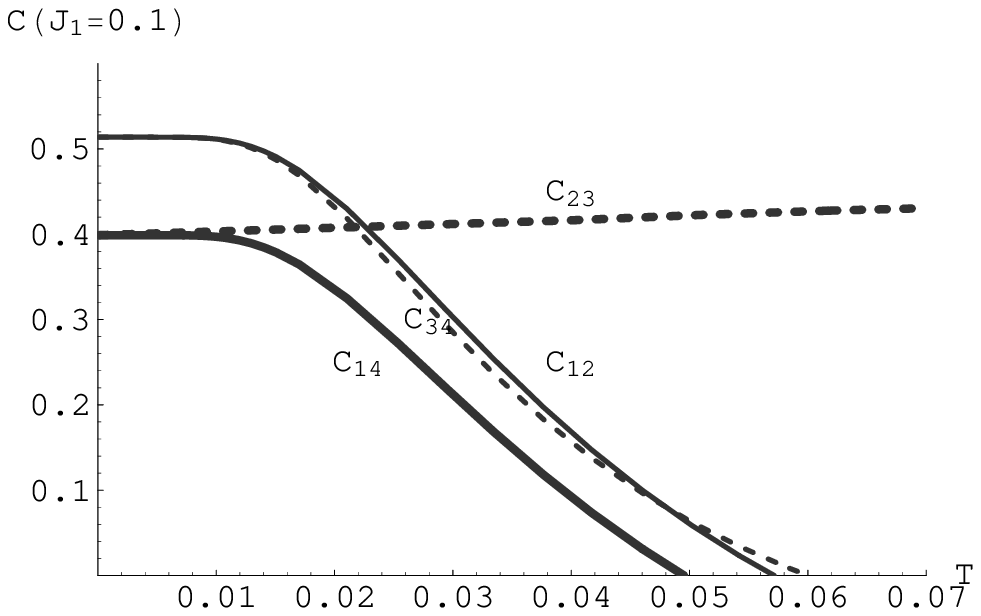}} {\epsfxsize
3cm \epsfysize 3cm \epsffile{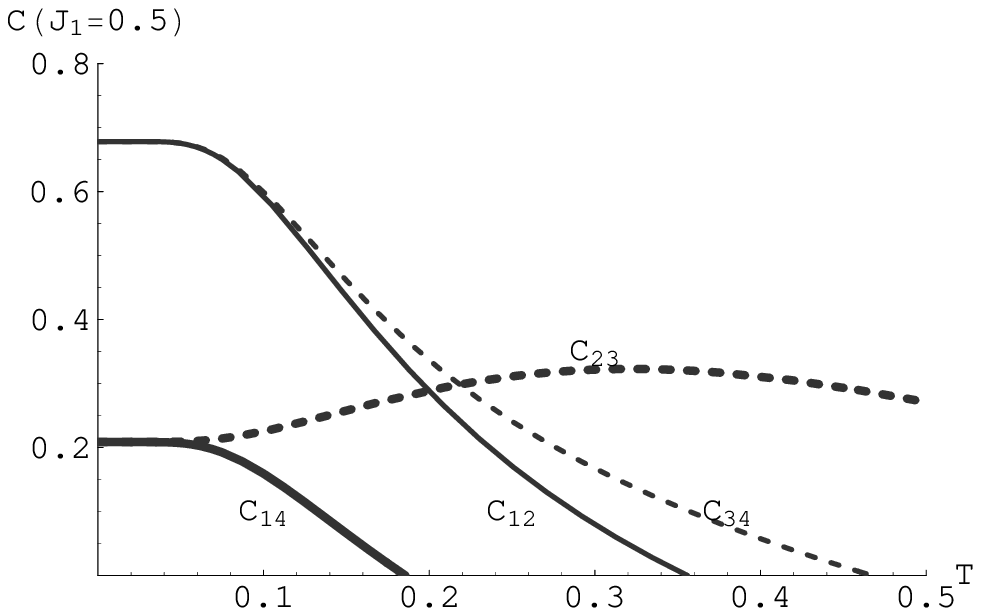}} {\epsfxsize 3cm
\epsfysize 3cm \epsffile{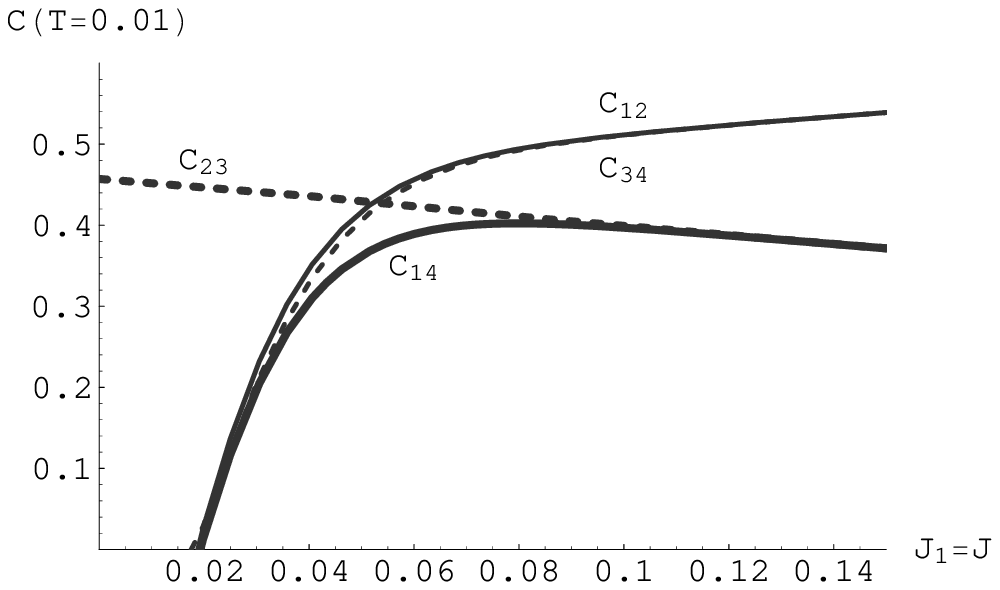}} {\epsfxsize 3cm \epsfysize
3cm \epsffile{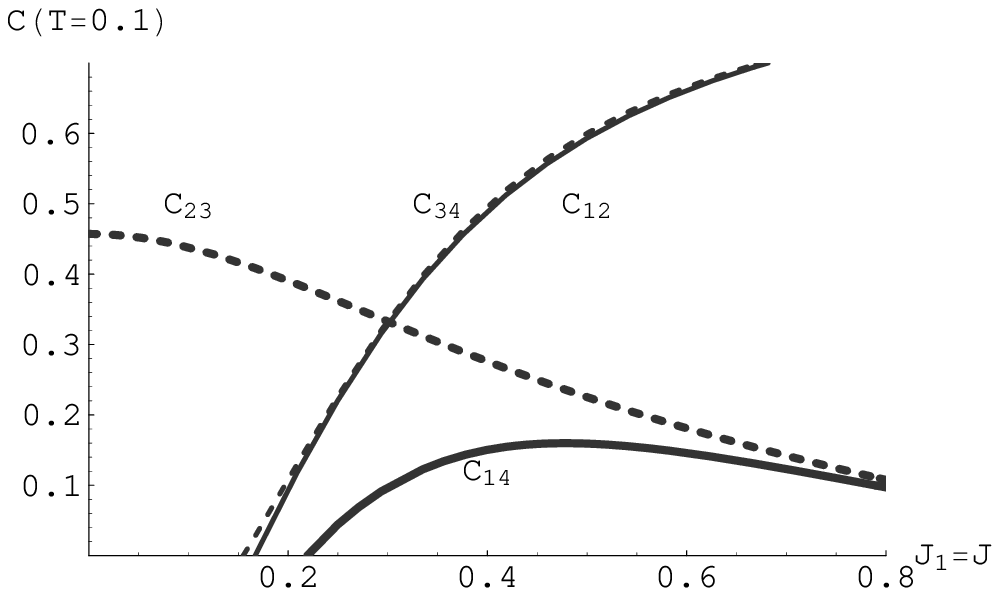}} \caption{The comparison among $C_{12},
C_{23}$, $C_{34}$ and $C_{14}$ at different fixed $J_1=J$ and
temperature.}
\end{figure}

From Fig. 4 we can see the transferred law of pairwise
entanglement: $C_{14}$ depends on the smallest of $C_{12},C_{34}$
and $C_{23}$ (min($C_{12},C_{34},C_{23}$)), when temperature is
lower, $C_{14}$ is equal to min($C_{12},C_{34},C_{23}$)
approximately, when temperature is higher, $C_{14}$ is less than
min($C_{12},C_{34},C_{23}$) but with similarly changed trend.
Changing the impurity site, similar results can be gotten. So if
we want to find the maximal $C_{14}$, we must increase
min($C_{12},C_{34},C_{23}$). Because $C_{23}$ is more important
than $C_{12}$ and $C_{34}$ so we call it "Entangled kernel".

We wander how $C_{14}$ change if making entangled kernel
$C_{23}=1$. Considering the nearest interaction theory (for
nearest pair, the stronger the interaction the bigger the pairwise
entanglement), we need to make the interaction between the second
and the third qubit strong, such a model is constructed in Eq.
(\ref{Ham}), let $N=4,J_1=J_4=J,J_2=J_3=1$, i.e the middle qubits
are normal sites and the boundary qubits are impurity sites, as
long as $J$ is smaller than 1, the above condition is satisfied,
obviously $C_{12}=C_{34}$ at this condition. Numerical results in
some condition are shown in Fig. 5.

\begin{figure}
{\epsfxsize 3cm \epsfysize 3cm \epsffile{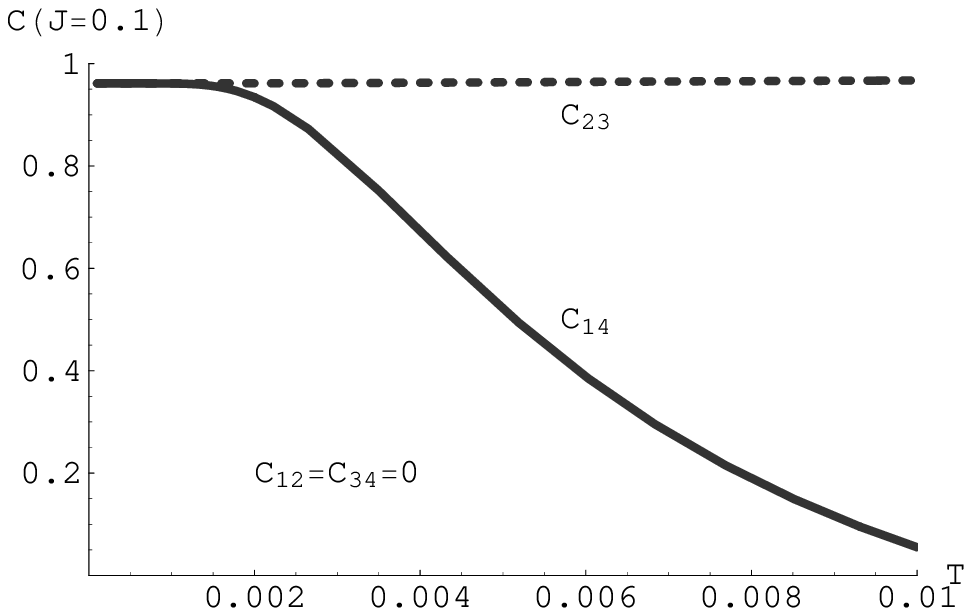}}
{\epsfxsize3cm \epsfysize 3cm \epsffile{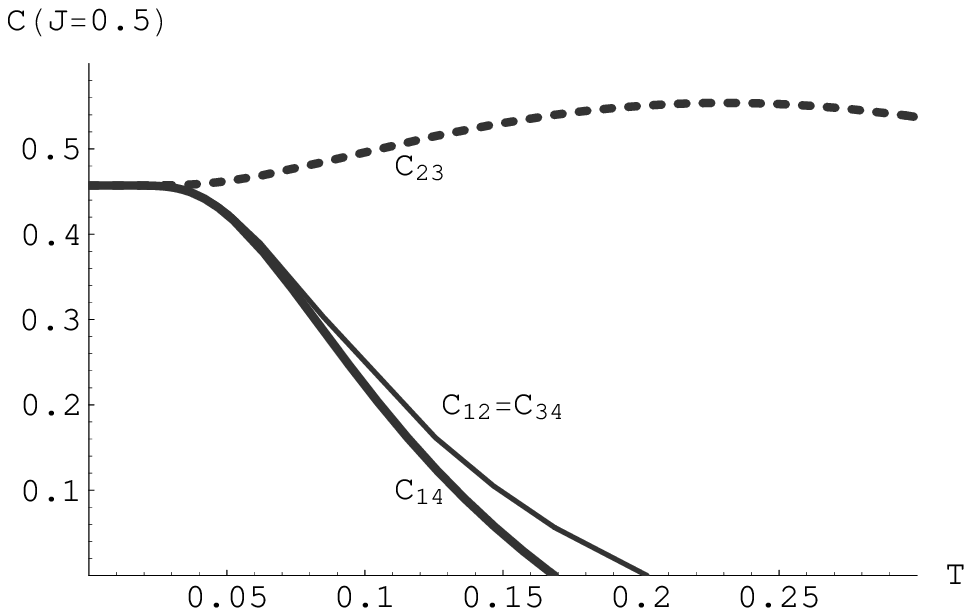}}
{\epsfxsize3cm \epsfysize 3cm \epsffile{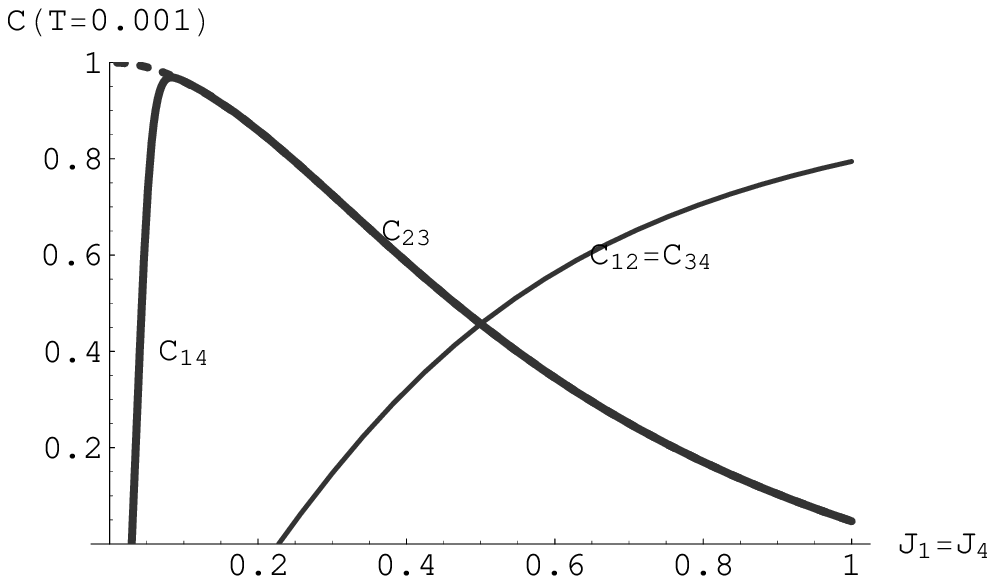}} {\epsfxsize
3cm \epsfysize 3cm \epsffile{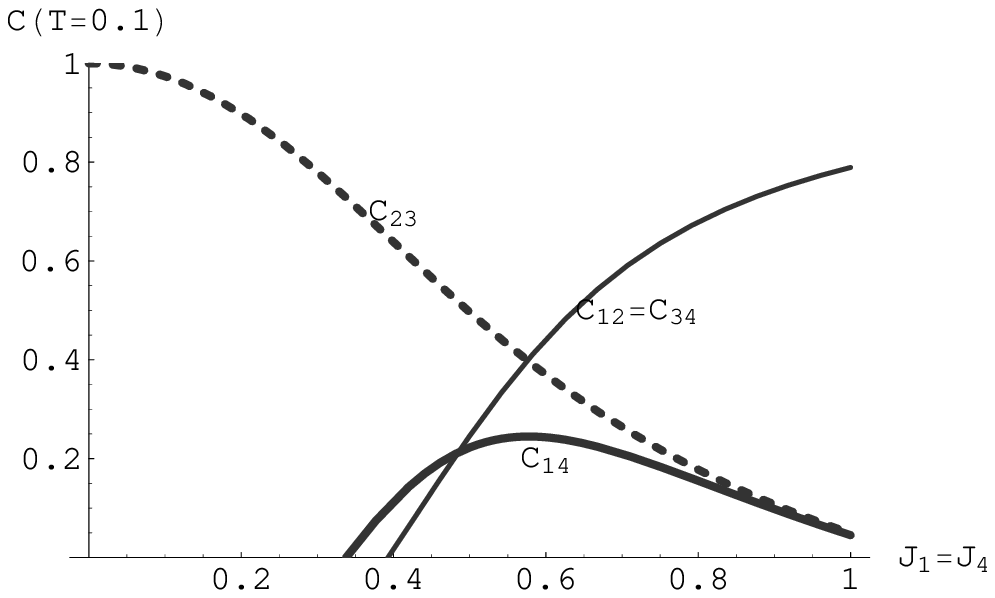}} \caption{The comparison
among $C_{12}, C_{23}$, $C_{34}$ and $C_{14}$ at different fixed
$J_1=J_4=J$ and fixed temperature.}
\end{figure}

Fig. 5 shows that at lower temperature, $C_{14}$ is only depended
on $C_{23}$ even $C_{12}=C_{34}=0$ (but their interaction can not
be too small). Choosing proper $J$ and temperature, we can make
$C_{14}=1$. This is an important support for the importance of
entangled kernel.

As a supporting for our theory, we calculate the pairwise
entanglement in the five-qubit open chain with boundary impurities
and find that $C_{13}=0, C_{15}=0$ and $C_{14} \ne 0$. For more
than five qubit case, the calculation will be very difficult and
tedious, we only construct a simple six-qubit case with
$J_1=J_6=0.1, J_2=J_5=1, J_3=J_4=10$, and get $C_{16}=0.96098$ at
the ground state(its entanglement is the maximal), of course we
can construct longer distance maximal pairwise entanglement as
long as the interaction strength decreases monotonously from
middle to boundary.

In this paper, we calculate all the pairwise entanglement in the
three and four-qubit Heisenberg open chain with system impurity,
and find some interesting results: for the nearest pairwise
entanglement, the stronger the interaction the bigger the
entanglement; for the non-nearest pairwise entanglement, the
condition of existing entanglement is decided by the qubit number
in the middle, odd qubit number means no entanglement, even qubit
number means the possibility of existing entanglement, this
conclusion is supported by $C_{13}=0$, $C_{15}=0$ and $C_{14} \ne
0$; the attenuation of non-nearest pairwise entanglement (if
exist) with temperature or impurity is quicker than the nearest
pairwise entanglement, i.e. the transferred entanglement is more
sensitive to the temperature and impurity.

The above conclusions are based on the open chain with qubit
number no more than six, but they have generality in other qubit
cases. First, pairwise entanglement is always connecting with
double qubits, so its transfer must be depended on the entangled
kernel, at this point they are coincidence; second, from the point
of physics, the transfer of pairwise entanglement is the transfer
of pairwise interaction, a best explanation is the example of
four-qubit with boundary impurities.

These conclusions are very important for the solid system, through
the transfer of entanglement we can construct pairwise
entanglement with longer distance, we can increase this distance
to micro-scale or even longer in theory, that is a practical scale
for solid state quantum information. As we know there have at
least two possible ways to realize such system: electrons floating
on liquid helium and electron spins in coupled semiconductor
quantum dots.

We don't know if the transfer of entanglement in spin chain is
still suitable for the entangled photons, because they have some
different in the essence of production. If it is possible, we will
no longer worry about the attenuation of entangled photons. We
also don't know if we can use the non-maximal pairwise
entanglement, if we can, at what value we can get the satisfied
fidelity in quantum information processing?

This work is supported by the NSF of China under grant 60490280,
90403034, 90406017,10547008 by the National Key Basic Research
Special Foundation of China under 2005CB724508, by the Foundation
of Xi'an Institute of Posts and Telecommunications under grant
105-0414, by the NSF of Shanxi Province under grant 2004A15.

\end{document}